\documentclass[12pt,preprint]{aastex}

\usepackage{lscape}

\newcommand{\spitzer}{{\it Spitzer~}}
\newcommand{\mic}{$\mu$m }

\shorttitle{GJ~436 Follow-up Photometry}
\shortauthors{Shporer et al.}

\begin{document}

\title{Photometric Follow-up Observations of the Transiting
Neptune-Mass Planet GJ~436b}

\author{
Avi Shporer\altaffilmark{1},
Tsevi Mazeh\altaffilmark{1},
Frederic Pont\altaffilmark{2},\\
Joshua N.\ Winn\altaffilmark{3},
Matthew J.\ Holman\altaffilmark{4},\\
David W.\ Latham\altaffilmark{4},
Gilbert A. Esquerdo\altaffilmark{4,5}
}

\altaffiltext{1}{Wise Observatory, Raymond and Beverly Sackler Faculty of Exact Sciences, Tel Aviv University, Tel
Aviv 69978, Israel}

\altaffiltext{2}{School of Physics, University of Exeter, Stocker Road, Exeter EX4 4QL, United Kingdom}

\altaffiltext{3}{Department of Physics, and Kavli Institute for Astrophysics and Space Research, Massachusetts Institute of Technology, Cambridge, MA, 02139, USA}

\altaffiltext{4}{Harvard-Smithsonian Center for Astrophysics, 60, Garden Street, Cambridge, MA, 02138, USA}

\altaffiltext{5}{Planetary Science Institute, 620 N. 6th Avenue, Tucson, Arizona 85705}

\begin{abstract}

This paper presents multi-band photometric follow-up observations of
the Neptune-mass transiting planet GJ~436b, consisting of $5$ new
ground-based transit light curves obtained in May 2007. Together
with one already published light curve we have at hand a total of $6$
light curves, spanning $29$
days. The analysis of the data yields an
orbital period $P=2.64386\pm0.00003$~days, mid-transit time $T_c$
[HJD] $=2454235.8355\pm0.0001$, planet mass $M_p =
23.1\pm0.9~M_{\earth} = 0.073\pm0.003~M_{Jup} $, planet radius
$R_p = 4.2\pm0.2~R_{\earth}=0.37\pm0.01~R_{Jup}$ and stellar
radius $R_s = 0.45\pm0.02~R_{\sun}$. Our typical precision for the
mid transit timing for each transit is about 30 seconds. We searched
the data for a possible signature of a second planet in the system
through transit timing variations (TTV) and variation of the impact
parameter. The analysis could not rule out a small, of the order of a
minute, TTV and a long-term modulation of the impact parameter,
of the order of $+0.2$ year$^{-1}$.

\end{abstract}

\keywords{stars: individual: GJ 436, planetary systems}

\section{Introduction}
\label{Introduction}

Of the almost $300$ extrasolar planets discovered to date\footnote{for
an updated list see: http://exoplanet.eu/}, the $\sim\! 35$ transiting
planets are the only ones allowing a measurement of their mass and radius
and the study of their atmospheres (e.g.,
\citealt{Charbonneau07}, \citealt{Fortney08}, \citealt{Guillot08},
\citealt{Pont08a}).
Of those, GJ~436b serves as a unique opportunity to study a planet
with mass and radius as small as Neptune. Although several planets
have already been detected with minimum masses similar to the mass of GJ~436b,
for example GJ~581b, c $\&$ d,
\citep{Bonfils05a, Udry07}, HD~4308b \citep{Udry06} and Cnc~55e
\citep{Fischer08}, GJ~436b is the only {\it transiting} Neptune-mass
planet discovered so far.  Moreover, it is the only known transiting
planet orbiting an M-type star, presenting an opportunity to help
constrain stellar models in this mass range.

GJ~436b was initially discovered by \citet{Butler04} through a radial
velocity (RV) modulation of its host star. \citet{Maness07} have
refined its orbital elements, specifically the eccentricity, $e =
0.160\pm0.019$, and identified a linear velocity trend of
$1.36\pm0.4$~m~s$^{-1}$~yr$^{-1}$. Those authors suggested that the
velocity trend might result from the presence of a long-period second
planet in the system, a planet that could induce a periodic modulation
of the orbital
eccentricity of GJ~436b. This type of effect was suggested already for
triple stellar systems \citep[][see also \citealt{Mazeh90}]{Mazeh79},
and for a planet in a binary system for 16~Cygni~B \citep{Mazeh97,
Holman97}.  \citet{Maness07} pointed out that this interpretation of
the eccentricity looks especially attractive if the circularization
timescale of GJ~436b is shorter than the age of the system, because
then we need to explain why the orbit of the planet has not been
circularized.

\citet{Butler04} obtained photometric observations of GJ~436 at the
expected time of possible transits and concluded that a transit was unlikely.
Nevertheless, the transiting nature of
GJ~436b was recently discovered by \citet{Gillon07a} who measured a
planetary radius of $\simeq\!  4~R_{\earth}$ and mass of $22.6~M_{\earth}$.
Soon after the discovery, a transit and secondary eclipse events were
observed by {\it Spitzer Space Telescope} at 8~\mic \citep{Gillon07b,
Deming07, Demory07}, further constraining the system parameters and reducing
the uncertainties by a
factor of about $10$ for the planet to star radius ratio and mid-transit
timing. Those observations resulted in a somewhat increased planetary radius,
of $4.2\pm0.2~R_{\earth}$.
\citet{Torres07} was able to refine further the planet
parameters by deriving more accurate constraints on the host star
and obtained a slightly larger planetary mass, $23.17\pm0.79~M_{\earth}$,
and a more precise planetary radius, of $4.22^{+0.09}_{-0.10}~R_{\earth}$.

Recently, \citet{Ribas08} restudied the system and suggested that the
observed radial velocities of the system are consistent with an additional small,
super-Earth planet in a close orbit around GJ~436 in a 1:2 mean-motion
resonance with the known planet. Similarly to \citet{Maness07}, they suggested
that this planet could pump eccentricity into the orbit.
\citet{Ribas08} further suggested that such an additional planet can
induce a precession of the orbital plane of GJ~436b, if the orbital
planes of motion of the two planets are inclined relative to each
other. Such a precession should change the inclination relative to our
line of sight and therefore might explain why \citet{Butler04} failed to
observe a transit in 2004.

Immediately after the detection of the transits of GJ~436b by
\citet{Gillon07a} we launched a ground-based (GB) observational campaign to
obtain high-quality transit light curves in different filters. Our
motivation was to obtain the first group of light curves which will be used
by future studies to look for a possible variation of the impact parameter
and to search for transit timing variations (TTV; \citealt{Agol05, Holman05}).
Our observations
are described in \S~\ref{Observations}. In our analysis, described in
\S~\ref{Analysis}, we simultaneously analyze six complete GB transit light curves: five
light curves obtained in our campaign and the light curve from
\citet{Gillon07a}.
In \S~\ref{Discussion} we
discuss our results.

\section{Observations}
\label{Observations}

In addition to the transit of UT~2007~May~2, which was reported by
\citet{Gillon07a} and was the first complete transit observation of
GJ~436b, we obtained five complete transit light curves of four
different transit events. The following paragraphs briefly describe
these observations.

We observed the transits of UT~2007~May~4 and May~25 with the Wise
Observatory 1~m telescope, located in Israel. We used a Princeton
Instruments (PI) VersArray camera and a $1340 \times 1300$ pixels
back-illuminated CCD, with a pixel scale of $0 \farcs 58$ per pixel,
giving a field of view of $13 \farcm 0 \times 12 \farcm 6$. On May 4
we used a Johnson $V$ filter and on May 25 a Bessel $R$ filter.  The
flexible scheduling of the Wise 1~m telescope was an important factor
in obtaining the transit light curve of May~4, less than three days
after the first observation of a complete transit.

The transit of UT~2007~May~4 was simultaneously observed with the Wise
Observatory 0.46~m telescope, operated remotely. The camera was a
$2148 \times 1472$ pixels Santa Barbara Instrument Group (SBIG) ST-10
XME CCD detector with a field of view of $40\farcm 5 \times 27 \farcm
3$ and a scale of $1 \farcs 1$ per pixel. This camera has no filters.
For a detailed description of this telescope and instrument see
\citet{Brosch08}.

On UT~2007~May~23 and May~31 we used the 1.2~m telescope at the
Fred L. Whipple Observatory (FLWO) on Mount Hopkins, Arizona.
The camera was the KeplerCam, which is a $4096^2$ Fairchild 486 CCD
with a square field of view $23 \farcm 1$ on a side \citep{Szentgyorgyi05}.
On both nights the filter was Sloan $z$ and a $2 \times 2$ binning was
used, giving a pixel scale of $0 \farcs 68$ per binned pixel.

Basic data reduction procedures, including bias and flat-field correction,
were applied to all images using standard IRAF\footnote{The Image Reduction
and Analysis Facility
(IRAF) is distributed by the National Optical Astronomy Observatory, which is
operated by the Association of Universities for Research in Astronomy, Inc.,
under cooperative agreement with the National Science Foundation.} routines.
We then performed aperture photometry of GJ 436 and several comparison stars of
similar brightness showing no significant variability.
The light curve of GJ 436
was calibrated by dividing it by the summed flux of the comparison stars.
Next, we fitted a polynomial of degree one or two to the out-of-transit (OOT) 
points vs.~time, and divided all points by this polynomial. The amplitude of
these polynomial corrections in mmag per transit duration was in the range of 
0.1--1.0 mmag.
As a final step, we divided the entire light curve by the OOT median intensity
and normalized the measurement uncertainties so the median
uncertainty OOT will equal the OOT RMS. Table~\ref{lclist} lists all
photometric measurements of the $5$ light curves obtained in this campaign.

Table~\ref{lctable} lists for each light curve the UT start date and time,
observatory and telescope, the filter used, limb darkening coefficients used 
in its analysis, average exposure time, average cadence (in min$^{-1}$), 
duration of the entire observation, start and end airmass, the $\beta$ 
factor (see below), RMS residuals, and the Photometric Noise Rate (PNR). 
The PNR is a quantity which takes into account the RMS and the cadence, 
defined as RMS$\times\sqrt{{\rm cadence}}$, in units of mmag per minute 
(see also \citealt[][section 2.2]{Burke08}). 

\section{Data Analysis and Results}
\label{Analysis}

We decided not to include the publicly available
\spitzer 8~\mic light curve in our simultaneous fitting since the shape of
that light curve is dependent on how the \spitzer ``ramp'' was modeled
\citep{Gillon07b, Deming07}. As the \spitzer measurements are much more precise
than the GB measurements, this ramp could have systematically influenced our results.
For comparison, the \spitzer light curve PNR is a factor of $\sim\!3$ smaller than 
that of our ground-based light curves.

Accounting for correlated noise \citep{Pont06} was done similarly to the
``time-averaging'' method of \cite{Winn08}. After a preliminary analysis
we binned the residual light curves
using bin sizes close to the duration of ingress and egress.
The presence of correlated noise in the data was quantified as the ratio
between the binned residual light curves standard deviation and
the expected standard deviation in the absence of correlated noise.
For each light curve we estimated $\beta$ as the largest ratio among
the bin sizes we used. We then multiplied the relative flux errors by
$\beta$. Values of $\beta$ are in the range of $1.0$--$2.1$ and
are listed in Table~\ref{lctable}.

\subsection{Simultaneous fitting of all parameters}
\label{Analysisall}

We fitted the system parameters to all six GB light curves simultaneously.
Our transit light curve model consisted of nine parameters:
The orbital period, $P$; a particular mid-transit time, $T_c$; planet to
star radius ratio, $R_p/R_s$; semimajor axis scaled by the
stellar radius, $a/R_s$; impact parameter, $b$ (See
Eq.~$3$ of \citealt{Winn07} for the formula of the impact parameter in
an eccentric orbit); two limb-darkening
coefficients $u_1$ and $u_2$, for a quadratic limb-darkening law;
orbital eccentricity, $e$; and longitude of periastron, $\omega$.
The latter four parameters were held fixed in the fitting process, hence our
model had $5$ free parameters.

For transiting planets on eccentric orbits the mid point, in time, 
between transit start and end is different than the time of 
sky-projected star-planet closest approach. For GJ~436b this difference is
comparable to the typical uncertainty on the mid-transit times obtained here. 
In our model the mid-transit time is defined as the time of closest
approach, which is also the time of minimum light.

For $e$ and $\omega$ we used the values
given by \citet{Maness07}. Although other authors (e.g., \citealt{Demory07},
\citealt{Deming07})
report more precise values
for $e$, their uncertainty on $\omega$ is either much higher than that of
\citet{Maness07} or not reported. As the errors in two orbital parameters
are correlated, we chose to adopt the reference giving the smallest
uncertainty on both.
To determine $u_1$ and $u_2$ we adopted a model of
$T_{\rm eff} = 3500$~K \citep{Maness07}, $\log g = 4.843$ \citep{Torres07} and
[Fe/H]=-0.03 \citep{Bonfils05b} and interpolated on
the \citet{Claret00,Claret04} grids.
For the light curve obtained with the Wise Observatory 0.46~m telescope,
which has no filter, we used limb-darkening coefficients corresponding to the
$R$ filter, as its CCD response resembles a ``wide-$R$''
filter \citep{Brosch08}. The coefficients used for each light curve
are listed in Table~\ref{lctable}.

We used the formulas of \citet{Mandel02} to determine the relative flux as
a function of the planet-star sky-projected separation and the
Monte Carlo Markov Chain
(MCMC) algorithm \citep[e.g.,][]{Tegmark04, Ford05, Gregory05}
to determine the parameters that best fit the
data, along with their uncertainties.
MCMC algorithms have already been used extensively
in the literature for fitting transit light curves
\citep[e.g.,][]{Holman06, Collier07, Gillon08, Burke07}.
Assuming a uniform prior for the five fitted parameters, the algorithm 
can be viewed as a random walk in the five-dimensional parameter space of 
the model.
Starting from an initial guess, at each step
the algorithm examines a new point in the parameter space which is reached
by adding a Gaussian random value to each of the parameters.
The algorithm decides whether to accept the new point in the parameter 
space and move there, or repeat the current point in the chain depending 
on the resulting posterior probability. If the new point has a higher 
relative posterior probability the algorithm will move there, and if not, 
it will move to the new point with a relative probability of
$\exp(-\Delta\chi^2/2)$, where $\Delta\chi^2$ is the $\chi^2$ 
difference between the two points in parameter space.
The distribution of parameter values in all points
is used to derive the best fit value and its uncertainties.
Here we took the distribution median to be the best fit value and the values at
the 84.13 and 15.87 percentiles to be the $+1\sigma$ and $-1\sigma$
confidence uncertainties, respectively.

The widths of the Gaussian distributions used to find a new point
are a crucial part of the MCMC algorithm and they affect the
fraction of accepted points, $f$. In order to have $f$ close to
$25\%$ \citep{Gregory05, Holman06} we
divided each long chain, of $500,000$ steps,
into smaller chains (mini-chains), of
$1,000$ steps, and re-evaluated the distribution
widths after the execution of each mini-chain. This re-evaluation
was done by scaling them according to the relative difference between $f$ of
the previous mini-chain, and the target value of $25\%$. The same scaling was
applied to all parameters.

To estimate the {\it relative} size of the distribution widths
for each one of the model parameters we initially performed a sequence
of $10$ mini-chains, where only that
parameter was allowed to vary and the target $f$ was $50\%$ \citep{Gregory05}.
Thus, a single MCMC run was comprised of several sequences of $10$
mini-chains and a long chain, of $500$ mini-chains. For determining the
final result we considered
only the long chain while ignoring the first $20\%$ of its steps.

We performed $10$ MCMC runs as described above while changing
each time the initial guess and initial distribution widths. Results of those
runs were highly consistent with each other and our final result is based
on all runs.

Before describing our results we briefly describe how we tested our analysis:

\begin{itemize}

\item To test the adaptive form of our MCMC algorithm we re-ran
it while fixing the distribution widths during the long chain part. 
The differences in the results for the fitted parameters did not 
exceed $0.15\sigma$, and the differences in the estimated uncertainties
were no more than $10\%$.

\item We repeated our analysis while skipping the polynomial corrections 
mentioned in Section~\ref{Observations}. The results for the fitted 
parameters were less than $0.2\sigma$ from the original ones, and the 
estimated uncertainties were similar at the $20\%$ level.

\item We applied our analysis on the \spitzer 8~\mic light curve alone, and derived
results similar to \citet{Gillon07b}, \citet{Deming07} and
\citet{Southworth08}, at the $\sim 1 \sigma$ level.

\item To further test the validity of our results we re-ran MCMC while adopting 
a Gaussian prior probability for $e$ and $\omega$. The central values and standard 
deviations were those given by \citet{Maness07}. 
The results were not modified significantly.

\end{itemize}

\begin{figure}
\includegraphics[scale=0.85]{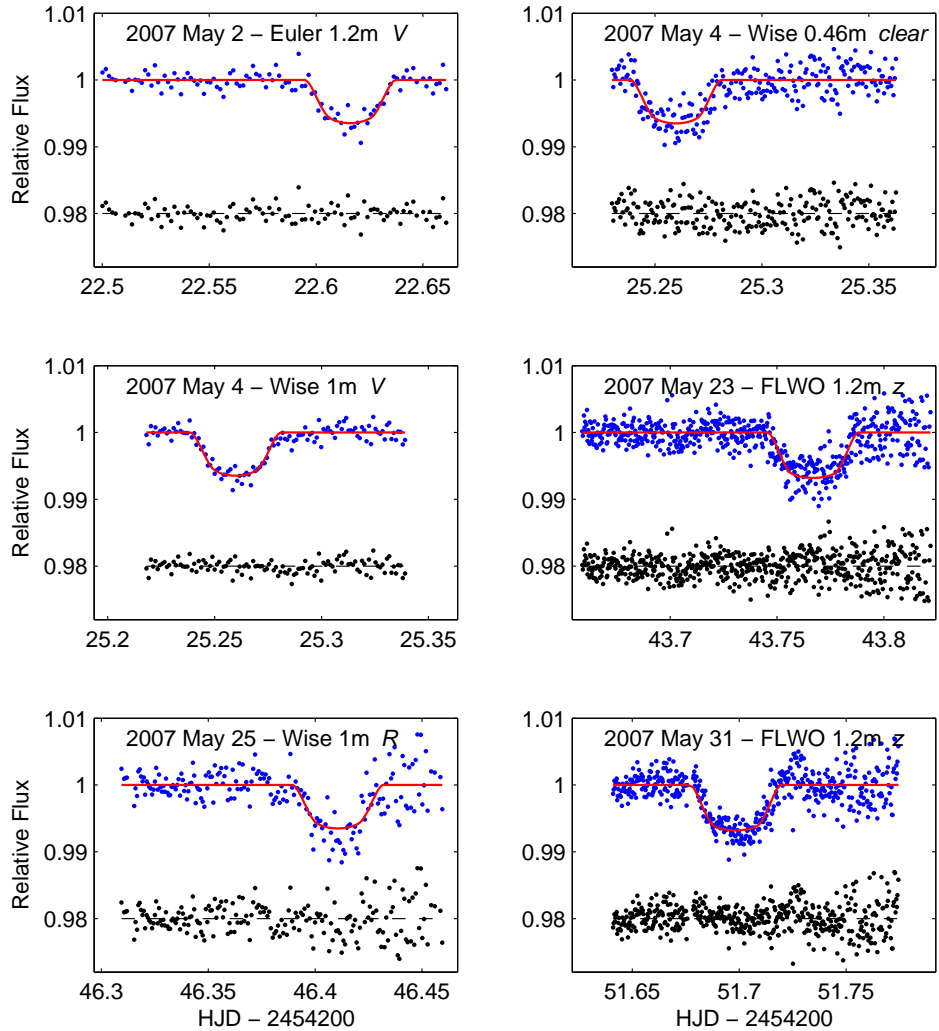}
\caption{
\label{lc}
The six light curves analyzed in this work. The overplotted solid
line is our best estimate model. For each light curve, residuals from
the model are plotted, centered on a relative flux of $0.98$ for clarity.
Within each panel the UT date, observatory, telescope and filter used
are given.}
\end{figure}

Fig.~\ref{lc} presents the six GB light curves, with our
fitted model overplotted in solid lines.
The three light curves obtained in the beginning of 2007~May
are of better quality, especially the Euler 1.2~m $V$ and Wise 1~m
$V$ light curves, with RMS residuals of $1.2$ and $0.9$ mmag,
respectively. The increased scatter in the three light curves
obtained at the end of 2007~May is the result of the high
airmass of these observations, done close to the end of the
observational season.

Our estimates for the fitted parameters, while
fitting all six light curves simultaneously,
are listed in Table~\ref{fitparams}. For comparison, this
table includes results from previous studies whenever these parameters were fitted
directly. In case they were not, but can be calculated
from other fitted parameters, they are given without an error.
\citet{Gillon07a} based their parameter determinations on an analysis
of the Euler 1.2~m $V$ light curve from UT~2007~May~2, which is included in this work.
The other three studies referred to in Table~\ref{fitparams} \citep{Gillon07b, Deming07,
Southworth08} are based on the \spitzer 8~\mic light curve. Results obtained by
\citet{Torres08} are not presented since for the light curve parameters of GJ~436b these
authors gave the weighted average of previous works.

Table~\ref{physparams} lists our results for the physical parameters
of the system, along with the corresponding values obtained previously.
Assuming the mass of the star to be $M_s=0.452\pm0.013 M_{\sun}$
\citep{Torres07} and adopting the orbital parameters of \citet{Maness07} (specifically:
$K$, $e$ and $\omega$) we derived the planet mass and semi-major axis.
Using Newton's version for Kepler's Third Law we derive the star and planet radius.

\subsection{Fitting transit timing and impact parameter for each light curve}
\label{Analysisind}

As mentioned in \S~\ref{Introduction}, \citet{Ribas08} suggested the presence
of a second planet orbiting GJ~436, in an outer orbit close to a 2:1 mean-motion
resonance with the transiting planet. The gravitational interaction between
the two planets may cause a detected TTV signal \citep{Holman05, Agol05}.
\citet{Ribas08} suggest that the second planet is responsible for changing
the inclination angle, $i$, of the transiting planet orbit. The presence of such
a planet may be detected through measuring transit timing and the impact parameter
of many transit light curves. As the light curves obtained here are the first
group of such light curves we performed two
additional MCMC analyses, described below.

First, we fitted $R_p/R_s$, $a/R_s$ and $b$ to all light curves
and an independent $T_c$ to each transit light curve, in order to estimate
the mid-transit time of each event. In this analysis the period
does {\it not} determine the time intervals between transit events. It
is used only for determining the true anomaly of each measurement within
each light curve so it is only weakly constrained. Hence, in this analysis $P$
was fixed at the value derived in our original analysis.
The resulting six $T_c$ values were later used to derive the transit ephemeris by a linear fit.
The new period is $\approx\! 0.35 \sigma$ from
the period fitted in our original MCMC run. Repeating the process described in this paragraph 
with the new period results in a variation of only $0.02\sigma$, so no further iterations 
were done.

Table~\ref{Tcb} contains the mid-transit timings fitted independently
to each light curve. Note that light curves with the smallest RMS residuals
do not necessarily have the smallest mid-transit time uncertainty. For example,
the FLWO 1.2 m light curve, at $E=3$, has relatively high RMS residuals although
a small mid-transit time uncertainty. This is simply the effect of the short
KeplerCam cadence, as the increased number of points act to better constrain
the mid-transit time. This effect is quantified by the PNR (see Table~\ref{lctable}).

By fitting a linear function,
\begin{equation}
T_c(E) = T_c(0) + P \times E,
\end{equation}
to the transit epoch, $E$, we derived $P=2.64386\pm0.00003$ days and
$T_c(0) = 2454235.8355\pm0.0001$ [HJD]. The transit event with $E=0$ was chosen
to be right in the middle of our observed events, although we did
not observe that
particular event. We take this ephemeris as our final
result, listed in Table~\ref{fitparams}, and note that it is consistent with
$P$ and $T_c$ derived from our original MCMC fitting.

The residuals from the linear fit are presented in Fig.~\ref{OC}, known as
the O-C diagram. The figure shows some indication
for a variability of the period, reflected by the high $\chi^2$
value of the fit, of $23$ for $4$ degrees of freedom.
However, the small number of points does not allow to assign any significance
to the detection of this variability, especially because some systematic
effects could shift some of the points.

Second, we performed a MCMC analysis where we looked for a transit
impact parameter variation.  The four parameters $P$, $T_c$, $R_p/R_s$
and $a/R_s$ were fitted to the entire data while the impact parameter,
$b$, was fitted independently to each light curve. The derived impact
parameters are listed in Table~\ref{Tcb} and plotted in Fig.~\ref{b}.
As a reference, we over-plotted in Fig.~\ref{b} the value of
our original run, in a solid line, and the upper and lower $1\sigma$
confidence limits in dashed lines.
No significant variation can be seen in the data, although a long-term
trend of the order of 0.2 year$^{-1}$ can not be ruled out.

The two separate analyses described in this section were repeated without 
applying the polynomial corrections mentioned in Section~\ref{Observations}.
The results were consistent with the original ones. 
In addition, for both the O-C transit timing residuals and the impact 
parameters derived here we looked for a possible correlation with several 
external parameters, including mid-transit airmass, filter central wavelength
and the light curves polynomial correction amplitude. 
All correlations were below the 1.5$\sigma$ confidence level.

\section{Discussion and Summary}
\label{Discussion}

This paper presents an analysis of $6$ complete transit
light curves of GJ~436b, spanning $29$ days, or $11$ orbital periods.

Our period is consistent with the period given by \citet{Maness07}, of
$2.64385\pm0.00009$ days, and is more precise.
Our value for $R_p/R_s$ (see Table~\ref{fitparams}), is larger,
by 1-2 $\sigma$, than the radii ratio derived for the \spitzer 8~\mic light
curve by \citet{Gillon07b}, \cite{Deming07} and \citet{Southworth08}.
If this difference is real it may be induced by stellar spots, or wavelength 
dependent opacities in the planetary atmosphere (\citealt{Barman07}, 
\citealt{Pont08a}, \citealt{Pont08b}). 
Interestingly, a similar effect, i.e.,
an increased radius ratio in the optical relative to the IR,
was already observed for HD~189733b \citep{Pont07}.

Our derived physical parameters, mainly the planetary radius and mass,
are consistent with those derived previously (See Table~\ref{physparams}).

\begin{figure}
\includegraphics[]{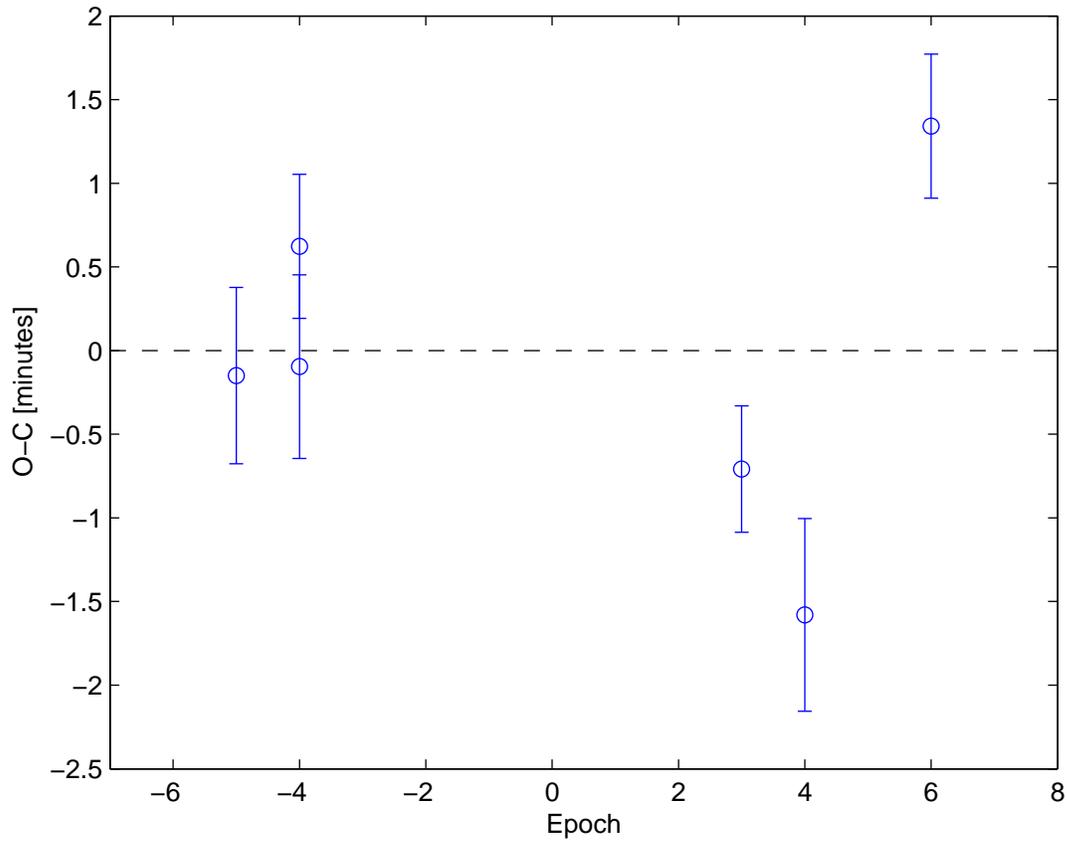}
\caption{
\label{OC}
Observed minus calculated (O-C) transit timing of the six light curves
included in this study. The graph shows the residuals from a linear fit
to the transit timing as a function of the transit epoch. Table~\ref{Tcb}
lists the actual transit timings.}
\end{figure}

\begin{figure}
\includegraphics[]{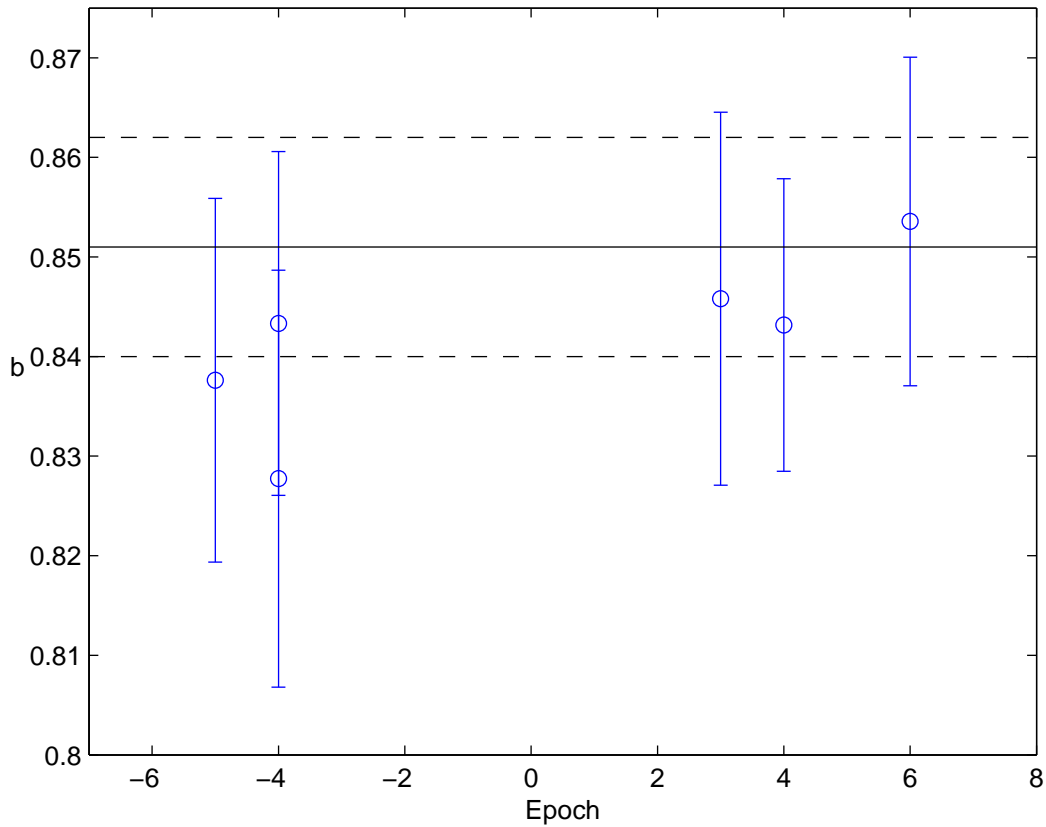}\\
\caption{
\label{b}
The impact parameter, $b$, fitted independently to each light curve, as a function
of the transit epoch.
The solid line marks the best estimate derived in our original MCMC run, while
fitting the impact parameter to all light curves simultaneously. The dashed lines
mark the upper and lower $1\sigma$ confidence limits. Table~\ref{Tcb} lists the actual
values of the impact parameters presented here.
}
\end{figure}

We examined the individual transit timing of each light
curve (TTV, see Fig.~\ref{OC}) and derived the impact parameter of
each light curve independently (see Fig.~\ref{b}). The transit timing
residuals show a possible hint for variability, supported by the
high $\chi^2$ of the fit. However, many more light curves are needed
to establish the variability and to explore its nature. The impact
parameter plot shows a possible trend, although a linear fit to the
six data points gives a slope consistent with zero.

Comparing our derived transit ephemeris to transit times published 
recently (\citealt{Alonso08}, \citealt{Bean08}) shows that the predicted
transit times are somewhat earlier than the measured times. The difference
is at the 1--2$\sigma$ level, so it is not highly significant. 
However, if real, it supports the claim made by \cite{Bean08} for a 
long-term drift in the transit times. Together with the 
long-term slope in GJ~436 RVs \citep{Maness07}
it suggests the existence of another object in the system, with a 
long period. Future
photometric and spectroscopic data will be able to better study
this possibility.

As mentioned in the introduction, the impact parameter of GJ~436b can
be changed by a precession of the line of nodes, induced by a second
planet whose orbital plane is inclined relative to the orbit of GJ~436b
\citep[e.g.,][]{Miralda02}. Such an effect was already suggested by
\citet{Soderhjelm75} and \citet{Mazeh76} for stellar triple systems.
The precession period is
\begin{equation}
\label{Unode}
U_{node}\approx\frac{M_s}{M_{p,2}}\left(\frac{P_2}{P}\right)^2P
\end{equation}
\citep{Miralda02}, where $M_{p,2}$ and $P_2$ are the mass and period
of the second planet, respectively. For the planet suggested by
\citet{Ribas08}, this precession is of the order of $10$ years.

For eccentric orbits the impact parameter can also be modified by
the precession of the periastron, because the impact parameter depends
on the orientation of the line of apsides (see
Eq.~$3$ of \citealt{Winn07}). The apsidal motion can be driven by the
stellar quadrupole moment, or by a second planet. The precession
driven by the stellar quadrupole moment could be quite slow, and
therefore probably does not contribute to the variation of the impact
parameter.  The rate of the apsidal precession induced by another
planet is of the same order of magnitude as the rate of the nodal
precession, estimated in Eq.~\ref{Unode}, and therefore can be of the
order of years, depending on the parameters of the unseen additional
planet in the system.

\begin{figure}
\includegraphics[]{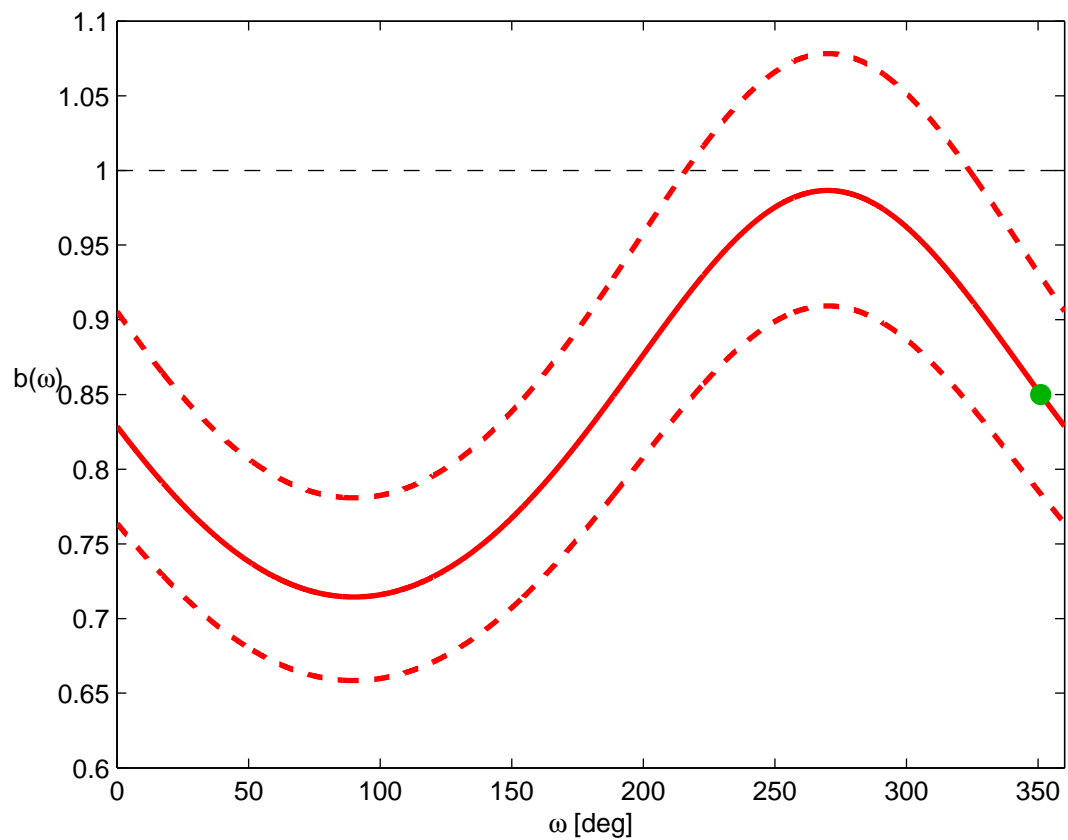}
\caption{
\label{bw}
The solid line presents the impact parameter, $b$, vs. $\omega$,
for the eccentricity and inclination of GJ~436b.
The filled circle (green) marks the current position, error bars are too small to be shown.
The upper and lower dashed lines (red) mark the impact parameter when the stellar
radius is multiplied by a factor of $1-R_p/R_s$ and $1+R_p/R_s$, respectively.
When multiplying the stellar radius by $1-R_p/R_s$ (upper dashed line)
the impact parameter will be smaller than one only for non-grazing transits.
When multiplying by $1+R_p/R_s$ (lower dashed line) then the impact parameter
will equal one when the minimal star-planet sky projected distance equals the
radii sum, i.e., for $b>1$ there will be no transit at all.
}
\end{figure}

In order to estimate the possible implication of the apsidal
precession on the impact parameter we plot in Fig.~\ref{bw} the impact
parameter of GJ~436b as a function of the argument of the periastron
$\omega$, at fixed orbital eccentricity and inclination.
We can see that $b$ is modulated between $\sim 1.0$ and
$0.7$ over the apsidal precession period. The upper dashed line of the
figure indicates that at some phase of the precession period the
transit of GJ~436b will even lose its flat-bottom shape. The figure
suggests that if the apsidal motion is of the order of ten years, we
should be able to observe soon a change of the impact parameter. In
case this modulation is due to apsidal motion, we will then be able to
confirm the change of the line of apsides direction by timing the
secondary eclipse or by precise RV measurements.

\acknowledgments

We are deeply thankful to Michael Gillon for his photometric processing 
the data obtained on 2007~May~4 and 2007~May~25. 
We thank the anonymous referee for his thorough reading of the paper 
and his useful comments which helped improve the paper.
TM and AS thank Elia Leibowitz and Liliana Formiggini for allowing the 
use of their telescope time at the Wise Observatory 1~m telescope on 
2007~May~4. 
This work was partly supported by Grant no.~$2006234$ from the United
States-Israel Binational Science Foundation (BSF), and by the Kepler
Mission under NASA Cooperative Agreement NCC 2-1390 with the Smithsonian
Astrophysical Observatory.



\clearpage


\begin{table}
\begin{center}
\caption{\label{lclist}
Photometry of GJ~436b. The table includes the $5$ light curves obtained in this work
and will appear in entirety in the electronic version of the journal.}
\begin{tabular}{lcccc}
\hline
\hline
Observatory+ & Filter & HJD             & Relative & Relative   \\
Telescope    &        &                 &  flux    & flux error \\
\hline
Wise 0.46m   & Clear  &  2454225.229864 & 1.0015        & 0.0020\\
Wise 1m      & V      &  2454225.218116 & 0.9997        & 0.0010\\
FLWO 1.2m    & z      &  2454243.658856 & 1.0002        & 0.0020\\
Wise 1m      & R      &  2454246.309414 & 1.0024        & 0.0017\\
FLWO 1.2m    & z      &  2454251.641074 & 0.9987        & 0.0024\\
\hline
\end{tabular}
\end{center}
\end{table}

\clearpage

\begin{deluxetable}{llrclccccccccccc}
\rotate
\tabletypesize{\scriptsize}
\tablecaption{\label{lctable}
List of light curves analyzed in this work.}
\tablewidth{22cm}
\tablehead{
\multicolumn{3}{c}{Start Date} & \colhead{Start Time} &
\colhead{Observatory+}& \colhead{Filter} & \colhead{$\ u_1$} & \colhead{$\ u_2$} & \colhead{exp.}& \colhead{Cadence} &
\colhead{Duration} & \colhead{start}& \colhead{end}& \colhead{$\beta$} & \colhead{RMS}  & \colhead{PNR\tablenotemark{a}}\\
\multicolumn{3}{c}{UT} & \colhead{[hh:mm] UT} & \colhead{Telescope} & & & &
\colhead{time [s]} & \colhead{[min$^{-1}$]}  & \colhead{[hours]}  & \colhead{AM} & \colhead{AM}&  & \colhead{[mmag]} & \colhead{[mmag min$^{-1}$]}}
\startdata
2007& May& 2    & 00:08      & Euler 1.2~m & $V$      & 0.340 & 0.444 & 80 & 0.44 & 3.86  & 2.10 & 2.24 & 1.0 & 1.2 & 1.8 \\
2007& May& 4    & 17:39      & Wise  0.46~m& $clear$  & 0.343 & 0.398 & 20 & 1.23 & 3.19  & 1.03 & 1.14 & 1.6 & 2.0 & 1.8 \\
2007& May& 4    & 17:22      & Wise  1~m   & $V$      & 0.340 & 0.444 & 60 & 0.65 & 2.90  & 1.05 & 1.07 & 1.5 & 0.9 & 1.1 \\
2007& May& 23   & 03:56      & FLWO  1.2~m & $z$      & 0.088 & 0.522 & 10 & 2.44 & 3.91  & 1.02 & 1.99 & 1.3 & 2.0 & 1.3 \\
2007& May& 25   & 19:33      & Wise  1~m   & $R$      & 0.343 & 0.398 & 40 & 0.88 & 3.60  & 1.15 & 3.79 & 1.3 & 2.4 & 2.6 \\
2007& May& 31   & 03:30      & FLWO  1.2~m & $z$      & 0.088 & 0.522 & 10 & 2.38 & 3.20  & 1.02 & 1.64 & 2.1 & 2.1 & 1.4 \\
\enddata
\tablenotetext{a}{Photometric Noise Rate = RMS$\times \sqrt{ {\rm Cadence}}$.}
\end{deluxetable}

\clearpage


\pagestyle{empty}
\begin{table}
\begin{center}
\caption{\label{fitparams}
Light curve fitted parameters.
}
\begin{tabular}{lr@{.}lr@{.}lr@{.}lr@{.}lr@{.}l}
\hline
\hline
Reference               & \multicolumn{2}{c}{$P$}   & \multicolumn{2}{c}{$T_c$-2454200}& \multicolumn{2}{c}{$R_p/R_s$}&\multicolumn{2}{c}{$a/R_s$} &\multicolumn{2}{c}{$b$}\\
                        & \multicolumn{2}{c}{[d]}   & \multicolumn{2}{c}{[HJD]}        & \multicolumn{2}{c}{}         &\multicolumn{2}{c}{}        &\multicolumn{2}{c}{}\\
\hline						  				 		    												
This work               &       $2$&$64386$         &   $35$&$8355$            &    $0$&$085$          &   $13$&$6$          &   $0$&$85$ \\
                        &    $\pm0$&$00003$         & $\pm0$&$0001$            & $\pm0$&$001$          & $\pm0$&$5$          &$\pm0$&$01$ \\
                        & \multicolumn{2}{c}{}      & \multicolumn{2}{c}{}     & \multicolumn{2}{c}{}  & \multicolumn{2}{c}{}&  \multicolumn{2}{c}{}\\
\hline
Gillon et al. 2007a$^1$ & \multicolumn{2}{c}{}      & \multicolumn{2}{c}{}     & $0$&$082$             & \multicolumn{2}{c}{} &\multicolumn{2}{c}{} \\
                        & \multicolumn{2}{c}{}      & \multicolumn{2}{c}{}     & $\pm0$&$005$          & \multicolumn{2}{c}{} &\multicolumn{2}{c}{} \\
\multicolumn{11}{c}{}\\
Gillon et al. 2007b$^1$ & \multicolumn{2}{c}{}      &   $80$&$78148$           & $0$&$0830$            & \multicolumn{2}{c}{} &  $0$&$849$\\
                        & \multicolumn{2}{c}{}      &   $+0$&$00015$           & \multicolumn{2}{c}{}  & \multicolumn{2}{c}{} &  $+0$&$010$\\
                        & \multicolumn{2}{c}{}      &   $-0$&$00008$           & \multicolumn{2}{c}{}  & \multicolumn{2}{c}{} &  $-0$&$013$\\
Deming et al. 2007$^1$  & \multicolumn{2}{c}{}      &   $80$&$78149$           & $0$&$0839$            &   $13$&$2$           &  $0$&$85$\\
                        & \multicolumn{2}{c}{}      & $\pm0$&$00016$           & $\pm0$&$0005$         & $\pm0$&$6$           &  $+0$&$03$\\
                        & \multicolumn{2}{c}{}      &  \multicolumn{2}{c}{}    & \multicolumn{2}{c}{}  & \multicolumn{2}{c}{} &  $-0$&$02$\\
Southworth 2008$^1$     & \multicolumn{2}{c}{}      &   $80$&$78174$           & $0$&$08284$           &   $13$&$68$          & \multicolumn{2}{c}{} \\
                        & \multicolumn{2}{c}{}      & $\pm0$&$00011$           & $\pm0$&$00090$        & $\pm0$&$51$          & \multicolumn{2}{c}{} \\
                        & \multicolumn{2}{c}{}      &  \multicolumn{2}{c}{}    & \multicolumn{2}{c}{}  & \multicolumn{2}{c}{} & \multicolumn{2}{c}{} \\
\hline
\end{tabular}
\end{center}
$^1$ Period was fixed at the Maness et al. 2007 value, of $P=2.64385\pm0.00009$ days.\\
\end{table}
\clearpage


\pagestyle{plain}
\begin{table}
\begin{center}
\caption{\label{physparams}
Results for the physical parameters.
}
\begin{tabular}{lr@{.}lr@{.}lr@{.}lr@{.}lr@{.}lr@{.}lr@{.}lr@{.}l}
\hline
\hline
Reference              &\multicolumn{4}{c}{$a$} &  \multicolumn{4}{c}{$M_p$}          &\multicolumn{4}{c}{$R_s$}       & \multicolumn{4}{c}{$R_p$} \\
                       &\multicolumn{4}{c}{[AU]}&  \multicolumn{4}{c}{[$M_{\earth}$]} &\multicolumn{4}{c}{[$R_{\sun}$]}& \multicolumn{4}{c}{[$R_{\earth}$]}\\
\hline
This work              &      $0$&$02872$   & $+0$&$00030$       &      $23$&$1$      &$\pm0$&$9$   &   $0$&$45$         &$\pm0$&$02$ & $4$&$2$  & $\pm0$&$2$\\
                       &\multicolumn{2}{l}{}& $-0$&$00025$       &\multicolumn{12}{l}{}\\
\hline
Gillon et al. 2007a    &\multicolumn{4}{l}{}                     &      $22$&$6$      &$\pm1$&$9$   &   $0$&$44$         &$\pm0$&$04$ & $3$&$95$ &   $+0$&$41$\\
                       &\multicolumn{4}{l}{}&\multicolumn{4}{l}{}&\multicolumn{6}{l}{}                                                           &   $-0$&$28$\\
Gillon et al. 2007b    &\multicolumn{4}{l}{}&\multicolumn{4}{l}{}                                   &   $0$&$463$        &  $+0$&$022$& $4$&$19$ &   $+0$&$21$\\
                       &\multicolumn{4}{l}{}&\multicolumn{4}{l}{}&\multicolumn{2}{l}{}&  $-0$&$017$ &\multicolumn{2}{l}{}                                 &   $-0$&$16$\\
Deming et al. 2007     &    $0$&$0291$      & $\pm0$&$0004$      &      $22$&$24$     &$\pm0$&$95$  &   $0$&$47$         &$\pm0$&$02$ & $4$&$33$ & $\pm0$&$18$\\
\multicolumn{17}{l}{}\\
Torres 2007            &    $0$&$02872$     & $\pm0$&$00027$     &     $23$&$17$      &$\pm0$&$79$  &   $0$&$464$        &  $+0$&$009$& $4$&$22$ &   $+0$&$09$\\
                       &\multicolumn{4}{l}{}&\multicolumn{4}{l}{}&\multicolumn{2}{l}{}                                   &  $-0$&$011$&\multicolumn{2}{l}{}&  $-0$&$10$\\
Southworth 2008$^1$    &\multicolumn{4}{l}{}&\multicolumn{4}{l}{}                                   &   $0$&$452$        &$\pm0$&$017$& $4$&$08$ &$\pm0$&$16$\\
 \multicolumn{17}{l}{} \\
\hline
\end{tabular}
\end{center}
$^1$Assuming $M_s=0.452^{+0.014}_{-0.012}~M_{\sun}$ (Torres 2007), orbital parameters of Maness et al.~(2007)
and the period derived here.\\
\end{table}
\clearpage


\pagestyle{plain}
\begin{table}
\begin{center}
\caption{\label{Tcb}
Mid-transit time and impact parameter fitted independently to each light curve$^\ast$.
}
\begin{tabular}{rr@{$\pm$}lr@{.}lr@{.}lr@{$\pm$}lr@{.}lr@{.}l}
\hline
\hline
E$\ $ & \multicolumn{2}{c}{$T_c$ - 2454200}& \multicolumn{4}{c}{$\Delta T_c$ $^\dagger$} & 
        \multicolumn{2}{c}{$b$}            & \multicolumn{4}{c}{$\Delta b$\ $^\ddagger$}   \\
      & \multicolumn{2}{c}{$\ \ \ $[HJD]}  & \multicolumn{2}{c}{[min]}        & \multicolumn{2}{c}{[$\sigma$]}   & 
        \multicolumn{2}{c}{}               & \multicolumn{2}{c}{}             & \multicolumn{2}{c}{[$\sigma$]} \\
\hline
-5 &   $22.61612$&$0.00037$ & -0&15 & -0&28 & $0.838$&$0.017$ & -0&013 & -0&61\\
			      	      	      		      	      	    
-4 &   $25.26002$&$0.00038$ & -0&10 & -0&18 & $0.828$&$0.020$ & -0&023 & -0&96\\
			      	      	      		      	      	    
-4 &   $25.26052$&$0.00030$ &  0&62 &  1&45 & $0.843$&$0.016$ & -0&008 & -0&37\\
			      	      	      		      	      	    
3  &   $43.76657$&$0.00026$ & -0&71 & -1&88 & $0.846$&$0.018$ & -0&005 & -0&23\\
			      	      	      		      	      	    
4  &   $46.40982$&$0.00040$ & -1&58 & -2&74 & $0.843$&$0.014$ & -0&008 & -0&41\\
			      	      	      		      	      	    
6  &   $51.69956$&$0.00030$ &  1&34 &  3&12 & $0.854$&$0.016$ &  0&003 &  0&13\\

\hline
\end{tabular}
\end{center}
$^\ast$ The table gives the result of two separate MCMC analyses.
In each analysis only a single parameter ($T_c$ or $b$) was fitted separately 
to every light curve and all other parameters were fitted to all light curves 
simultaneously, except for $P$ which was fixed when fitting an individual $T_c$ 
to each light curve.\\
$^\dagger$ Difference between the measured $T_c$ and the expected mid-transit 
time according to the fitted ephemeris (see Table~\ref{fitparams}).   
The difference is given in minutes and also in units of the uncertainty
on each mid-transit time. \\
$^\ddagger$ Difference between the measured impact parameter while fitting 
each light curve separately, and the result of the simultaneous fit (see 
Table~\ref{fitparams}). The difference is given also in units of the 
uncertainty on each impact parameter. 
\end{table}

\clearpage

\clearpage

\clearpage

\clearpage

\clearpage

\end{document}